\documentclass[floatfix,showpacs,showkeys,amsmath,amssymb,aps,pra,twocolumn,10pt]{revtex4-1} %

\usepackage[USenglish]{babel} 
\usepackage{overpic}
\usepackage{dsfont}
\usepackage{rotating}
\usepackage{xcolor}
\usepackage[urlcolor=blue,colorlinks=true,linkcolor=blue,pdfstartview={FitH},bookmarks=false]{hyperref} 
\newcommand{\ket}[1]{\ensuremath{|#1\rangle}}
\newcommand{\bra}[1]{\ensuremath{\langle#1|}}
\newcommand{\iner}[2]{\braket{#1}{#2}}
\newcommand{\braket}[2]{\ensuremath{\langle#1|#2\rangle}}
\newcommand{\eg}{\emph{e.g.}}
\newcommand{\ie}{\emph{i.e.}}
\newcommand{\etal}{\emph{et al.}}

\newcommand{\smal}{\scriptscriptstyle}
\newcommand{\ot}{\otimes}
\newcommand{\op}{\oplus}
\newcommand{\mc}{\mathcal}
\newcommand{\mb}{\mathbb}
\newcommand{\dg}{\dagger}
\newcommand{\idQ}{\mathbb{I}_{Q}}
\newcommand{\idE}{\mathbb{I}_{E}}
\newcommand{\bs}{\boldsymbol}

\begin{document}

\title{Exact reduced dynamics for a qubit in a precessing magnetic field and in the contact with a heat-bath}             

\date{26 May 2010}

\author{Bart\l{}omiej Gardas}
\email{bartek.gardas@gmail.com}
\affiliation{Institute of Theoretical and Applied Informatics, Polish Academy
of Sciences, Ba{\l}tycka 5, 44-100 Gliwice, Poland}

\begin{abstract}
The two-level quantum system (qubit) in a precessing magnetic field and
in contact with a heat bath is investigated. The exact reduced dynamics for the
qubit in question is obtained. We apply the approach based on the block operator
matrices theory, in contrast with the standard methods provided by the theory of
the open quantum systems. We also present the solution of the Riccati operator 
equation associated with the Hamiltonian of the system. Next, we study the adiabatic
approximation for the system under consideration using quantum fidelity as a way to
measure validity of the adiabatic theory. We find that in the weak coupling domain
the standard condition that ensures adiabatic behavior of the spin in the magnetic
field also guarantees its adiabatic evolution in the open system variant of this 
model. Therefore, we provide the explicit example of the open quantum system that
satisfies the adiabatic approximation firstly formulated for the closed quantum systems.
\end{abstract}
\maketitle



\section{Introduction} \label{sec:intro}

The most established and useful time-dependent two-level quantum system,
is perhaps, the one that describes spin half particle (qubit $Q$) in a
precessing magnetic field~\cite{galindo2}. The Hamiltonian of this system
in its basic variant is usually written in the following form

\begin{equation}
 \label{spin}
      H_{Q}(t)= -\gamma\mathbf{S}\cdot\mathbf{B}(t), 
      \quad \mathbf{S}=\boldsymbol{\sigma}/2, 
\end{equation}
where $\gamma$ is a constant, called the gyromagnetic ratio (the specific
value of $\gamma$ is irrelevant in context of our discussion), $\boldsymbol{
\sigma}\equiv(\sigma^x,\sigma^y,\sigma^z)$ and $\sigma^i$, $(i=x,y,z)$ are
the standard Pauli matrices. The magnetic field $\mathbf{B}(t)$ is assumed
to have the form

\begin{equation}
 \label{field}
       \mathbf{B}(t) = B_1\cos\left(\omega t\right)\hat{e}_x+ 
        B_1\sin\left(\omega t\right)\hat{e}_y+B_0\hat{e}_z, 
 \end{equation}
where $\hat{e}_x$, $\hat{e}_y$, $\hat{e}_z$ are the unit vectors along $x$,
$y$ and $z$ axis, respectively. $B_0$, $B_1$ define the amplitudes of the
magnetic field and $\omega$ is the frequency of the rotation. Using the 
equations above one can also rewrite the Hamiltonian~(\ref{spin}) in a more
readable form:

\begin{equation}
 \label{spin2}
      H_{Q}(t,\beta) = \beta\sigma_3+\alpha\left(\sigma_2\sin\left(\omega t\right)
      +\sigma_1\cos\left(\omega t\right)\right), 
\end{equation} 
where for the sake of convenience the abbreviations $\alpha =\tfrac{1}{2}
\omega_1\equiv-\tfrac{1}{2}\gamma B_1$ and $\beta = \tfrac{1}{2}\omega_0\equiv
-\tfrac{1}{2}\gamma B_0$ were introduced.
 
In the case of no coupling with the external environment (heat-bath) the exact
form of the evolution operator $U_{Q}(t)$ and hence the density matrix 
$\rho_Q(t)=U_{Q}(t) \rho_{Q}U_{Q}(t)^{\dagger}$ for the model~(\ref{spin2}), 
can be derived in an elegant and simple manner (see \eg~\cite{thaller,galindo2}).
This problem is so common that it can be found in almost every
modern textbook on quantum mechanics. However, if the aforementioned coupling 
is present, the \emph{exact} form of the density matrix has not yet been derived.

If one allows the system to interact (not necessarily to exchange the energy)
with the environment then it becomes so called \emph{open quantum system}
~\cite{breuer}. Its time evolution is \emph{not} unitary any more because of 
the decoherence process~\cite{Leggett,zurek}. Nevertheless, it may always be 
described by trace preserving (TP) and completely positive (CP) quantum operation 
(or quantum channel, see \eg~\cite{geometry}) $T_{t}: \rho_Q\rightarrow\rho_{Q}(t)$,
 $\rho_{Q}(0)\equiv \rho_{Q}$. Unfortunately, finding its exact form is almost 
impossible in most cases, especially for the systems described by the time-dependent
Hamiltonian.
                                                                      
Naturally the question arises: how difficult this problem actually is?
Recently, the efforts to answer this question were made ~\cite{mgr,gardas}.
It was shown that from the mathematical perspective this task is at least as
complicated as the problem of resolving the Riccati algebraic equation associated
with the Hamiltonian defining the model in question. Moreover, from the block 
operator matrices theory point of view obtaining the exact reduced dynamics for any
system where decoherence is present (beyond dephasing phenomenon) ~\cite{dajka,dajka2}
is as complex as resolving this equation. Of course, the complexity of the analysis
strongly depends on the particular choice of the environment responsible for the 
decoherence process.
                                                                                                 
First, we show that all the difficulties mentioned above, including the problem of
resolving the Riccati equation, can be overcome for the so-called spin environment.
Therefore, we provide the exact reduced dynamics for the open system version of the
model~(\ref{spin}) in the case when the system is immersed within the spin-bath. Next,
as an application, we investigate the adiabatic approximation for the open system
constructed in this way. Moreover, we show that in the weak coupling domain the 
condition ensuring the adiabatic evolution of the system~(\ref{spin}) also leads to 
the adiabatic behavior of its generalization.

The layout of this paper is as follows. In Sec.~(\ref{matrix}) we briefly review the
concept of the block operator matrix and the procedure allowing one to 
diagonalize it. In Sec.~(\ref{sub:partial}) we introduce the Riccati operator
equation and discuss its connection with the Hamiltonian specifying the model.
In Sec.~(\ref{sec:model}) the definition of the model is given and the exact
reduced dynamics is derived. We also indicate possible applications of the model.
In Sec.~(\ref{sec:adiab}) we investigate the adiabatic approximation. Finally, 
Sec.~(\ref{sec:summary}) contains the final remarks and the summary of the paper.
  

\section{Block operator matrices approach}
   \label{matrix}

  \subsection{Partial trace and reduced dynamics}
     \label{sub:partial}
  
Before we can go any further and present the main result of this paper let us
first discuss the procedure allowing one to obtain the density matrix $\rho_Q(t)$
for a two-dimensional system (qubit) using the block operator matrices approach. 

Let $H_{QE}$ be the Hamiltonian of the closed (qubit and environment, $Q+E$)
system. We will assume its following form

\begin{equation}
  \label{qe}
        H_{QE} = H_{Q}\ot \mb{I}_{E}+\mb{I}_{Q}\ot H_{E}+H_{int}, 
\end{equation}
where $H_Q$, $H_E$ are the Hamiltonian of the qubit and the environment
respectively, while $H_{int}$ represents the interaction between the systems.
Hamiltonian~(\ref{qe}) acts on $\mc{H}_{QE}= \mc{H}_{Q}\ot\mc{H}_{E}$ space,
where $\mc{H}_{Q}$ and $\mc{H}_{E}$ are the Hilbert spaces for the qubit and the
environment, respectively. The state $\rho_{Q}(t)$ of the qubit at any given time
$t$ can be computed using the following formula (for more details see \eg~\cite{alicki} )

\begin{equation} 
 \label{trace}
   \begin{split}
        \rho_{Q}(t) &=\mbox{Tr}_{E}(U_t\rho_{Q}\otimes\rho_{E} U_t^{\dagger})\\
                    &\equiv T_t(\rho_{Q}),
  \end{split}
\end{equation}
where $U_t=\exp (-iH_{QE}t)$ is the evolution operator of the total $Q+E$ system
(we work with units $\hbar=1$). By $\mbox{Tr}_{E}(\cdot)$ we denoted the
partial trace. In the literature the quantity $\rho_Q(t)$ is commonly called the \emph{reduced
dynamics}. In this paper we will be also referring to it as the solution of a given
model. In Eq.~(\ref{trace}) the assumption that $\rho_{QE}=\rho_{Q}\ot\rho_{E}$ 
was made, \ie, no correlation between $Q$ and $E$ are present initially~(see
\eg, \cite{korelacje,*erratum,KrausRep} and references therein).

Since $\mc{H}_{Q}=\mb{C}^2$ and the isomorphism $\mb{C}^2\ot\mc{H}_{E}=\mc{H}_{E}\op\mc{H}_{E}$
holds, one can think of the operator $H_{QE}$ (and of any given
operator $A$ acting on $\mc{H}_{Q}\ot\mc{H}_{E}$) as a $2\times2$ block operator
matrix~\cite{bom}. Let us assume that $A$ acts on $\mc{H}_{E}\op\mc{H}_{E}$.
Thus, we can write $A=[A_{ij}]$, where operators $A_{ij}$, $(i,j=1,2)$ act on
$\mc{H}_{E}$. In this block operator matrices nomenclature the procedure of calculating the
partial trace $\mbox{Tr}_{E}(\cdot)$ can be written as 

\begin{equation}
 \label{partial}
   \mbox{Tr}_{E}(A)=
     \left( 
       \begin{array}{cc}
         \mbox{Tr} A_{11} & \mbox{Tr}A_{12}   \\
         \mbox{Tr}A_{21} & \mbox{Tr}A_{22} 
       \end{array}
     \right),
\end{equation}
where $\mbox{Tr}(\cdot)$ denotes the trace on $\mc{H}_{Q}$. Eq.~(\ref{partial})
introduces the concept of the partial trace in a very simple and, what is the
most important, intuitive way. It is also possible to define the partial
trace locally, \ie, $\mbox{Tr}_{E}(X\ot Y) =X\mbox{Tr}(Y)$, where as mentioned before
$\mbox{Tr}(\cdot)$ stands for the trace operation on the space on which the
operator $Y$ acts on. From Eq.~(\ref{partial}) one can learn that the partial trace
is the operation which transforms operator matrices into the ``ordinary''
matrices. Moreover, the partial trace is a linear operation going from
$B(\mc{H}_{QE})$ space to the $B(\mb{C}^2)$ space. Furthermore, from the block operator matrices
perspective one can easily understand why dealing with the open systems is much
more complex (from the mathematical viewpoint) then dealing with the closed
systems. The reason is that the procedure allowing one to calculate the reduced
dynamics is not as straightforward as Eq.~(\ref{partial}) might indicate.
Indeed, to compute the partial trace~(\ref{trace}) one needs to write the
evolution operator $U_t$ in a $2\times 2$ block matrix form. The last operation
requires diagonalization of the block operator matrix $H_{QE}$, which is much
more complicated than the diagonalization of the standard $2\times 2$ matrix. It
leads to the problem with solving the Riccati algebraic equation analyzed below.

\subsection{Riccati equation and block diagonalization}
 \label{sec:riccati} 

With every Hermitian block operator matrix of the form 

\begin{equation}
 \label{bom}
   R=
     \begin{bmatrix}
       A & B \\
       B^{\dg} & C
     \end{bmatrix},
\end{equation}
where $A$, $B$, $C$ are the Hermitian operators acting on $\mc{H}_{E}$, one can 
associate the following Riccati equation~\cite{RiccEq}

\begin{equation}
 \label{Riccati}
   XBX+XA-CX-B^{\dg}=0,
\end{equation}
where $X\in B(\mc{H}_{E})$ is the operator to be determined. 

If the solution $X$ of the Eq.~(\ref{Riccati}) exists, then it may be used to
diagonalize operator matrix~(\ref{bom}). The procedure that allows us to do so
is as follows (for details see~\cite{gardas} and references therein)

\begin{equation}
 \label{xdiag}
U_{X}^{-1}RU_{X}=
   \begin{bmatrix}
     A+BX  & 0_{E} \\
     0_{E} & C-(XB)^{\dg}
    \end{bmatrix},
\end{equation}
where the similarity matrix $U_{X}$ is given by

\begin{equation}
 \label{ux}
U_{X}=
   \begin{bmatrix}
     \idE & -X^{\dg} \\
      X   & \idE
    \end{bmatrix}.
\end{equation}
We wish to emphasize that in most cases a general method of finding the solution
for the Riccati equation does not exist. Therefore, resolving the Eq.~(\ref{Riccati})
with respect to $X$ is a very difficult task. Moreover, even if the solution is known,
 there is still a problem of computing the inversed operator matrix $(U_{X})^{-1}$. 
Formally, if the operator $X$ is normal, \ie, $[X,X^{\dagger}]=0$, then this matrix
is given by the formula

\begin{equation}
 \label{Xinverse}
 U_{X}^{-1} = (\idE+XX^{\dagger})^{-1}
    \begin{bmatrix}
      \idE & X^{\dg} \\
       -X   & \idE
\end{bmatrix}.
\end{equation}
Otherwise, the form of the inverse operator $U_{X}^{-1}$ becomes more complicated.
Fortunately, the Eq.~(\ref{Xinverse}) is sufficient for our analysis. Notice that
it may be difficult to handle computationally the operator such as $(\idE+XX^{\dagger})^{-1}$.
In general the operator $X$ does not need to be Hermitian or unitary.
 
Perhaps those are the reasons why the approach to open systems based on the
operator matrices theory has not attracted too much attention. In our opinion,
however, the operator matrix perspective can offers a better understanding of
the open quantum systems and might give some new results in this area.
    

\section{Exact reduced dynamics}
 \label{sec:model}

\subsection{Model}
  \label{model}
 
We consider the model defined by the following time-dependent Hamiltonian

\begin{equation}
 \label{ht} 
     H_{QE}(t,\beta) = H_{Q}(t,\beta)\ot\mb{I}_{E}+
                       \mb{I}_{Q}\ot H_{E}+H_{int}, 
\end{equation}
where $H_{Q}(t,\beta)$ is given by Eq.~(\ref{spin2}) and  it represents a
qubit in the magnetic field~(\ref{field}). The environment is composed of $N$
independent and non-interacting spin $1/2$ particles. The Hamiltonian $H_{E}$
of the bath is assumed to be of the form 

\begin{equation}
 \label{hr}
    H_{E}= \sum_{n=1}^{N}\omega_n\sigma_n^{z}, 
\end{equation}
where $\omega_n$ and $\sigma_n^z$, $(n=1,...,N)$ are the frequencies and the Pauli
matrices for $n$th qubit, respectively. The Hilbert space $\mc{H}_{E}$ on which
the Hamiltonian~(\ref{hr}) acts is given by $N$-fold tensor product of $\mb{C}^2$ 
spaces, \ie, $\mc{H}=\overset{N}{\underset{n=1}{\bigotimes}}\mb{C}^2$. Therefore
for any $n\leq N$ the operator $\sigma_n^z$ is understood as $\sigma_n^z=
I_2\ot\dots\ot\sigma^z\ot\dots\ot I_2$, where $\sigma^z$ is the standard Pauli 
matrix acting on $\mb{C}^2$ and $I_2$ is the $2\times 2$ identity matrix on that 
space. In our model the coupling of a qubit with the environment is governed by 
the Ising-typ Hamiltonian~\cite{spinStar1,spinStar2}, namely

\begin{equation}
\label{ising}
   H_{int}=\sigma^z\ot\sum_{n=1}^{N}g_n\sigma_n^z, 
\end{equation}
where $g_n$ are the coupling constants. We assume that the bath is initially in
the Gibbs thermal state at a temperature $T$, \ie,

\begin{equation}
 \label{gibbs}
    \rho_{E}= Z^{-1}\exp(-H_{E}/kT), 
\end{equation}
where $Z=\mbox{Tr}(\exp(-H_{E}/kT))$ and $k$ is the Boltzmann constant.

The model described by the Hamiltonian~(\ref{ht}) with $H_E$ and $V$ specified 
by the Eqs.~(\ref{hr})-(\ref{ising}) in the case of the $\alpha=0$ (\ie, dephasing
and static magnetic field case) was investigated both in the context of the approximation
methods in the open quantum systems theory~\cite{spinStar1} and the capacities of
the quantum channels~\cite{spinStar2}. Thus, for the detailed discussion and possible
applications we refer the reader therein.

The quantum system discussed here can by thought of as the generalization of the model 
introduced in the mentioned papers to the case when the energy is exchanged between the
systems and the magnetic field depends on time. It is worth mentioning physical problems
that make use of this model. First of all, the Hamiltonian~(\ref{ht}) 
may pose a useful prototype describing any spin ``resonance'' phenomenon, like for instance
Nuclear Magnetic Resonance (NMR)~\cite{Kittel,NMR}. In such picture the spin $1/2$ particle
is the open system, which evolution in time we wish to describe. The spin-bath models the
influence of the other spins on the open system in question. Finally, the rotating magnetic
field is used to induce the resonance and to control the system.

The quantum devices of the future, like for instance quantum computers~\cite{QCNature} 
will be probably built from the components that are composed of a large amount of the qubits (spins,
in particular) \eg, quantum memory~\cite{QMNature}, quantum register~\cite{Register2,*{QRNature}},
etc. The model we are investigating may serves as a simple prototype allowing one to trace the
evolution in time of a single qubit of the register or memory. The magnetic field may be then
applied to program the device. Those are just a few potential applications of the simple
theoretical, time-dependent spin-spins model~(\ref{ht}).


\subsection{Exact solution}

To derive an exact reduced dynamics of the system governed by the Hamiltonian
~(\ref{ht}) we first simplify this problem to a time-independent one. In order
to accomplish this we use recently proved theorem that says if (for more details
see~(\ref{sec:app}) and also~\cite{gardas})

 \begin{enumerate}
  \item[i)]   the total system is defined by the Hamiltonian~(\ref{ht}),
  \item[ii)]  the interaction term $H_{int}$ between the systems takes the 
                  form $f(\sigma^z)\ot V$, where $f$ is an analytical
                  function of $\sigma^z$ and $V$ is a Hermitian operator,           
  \item[iii)]    $\rho_t(\beta)$ represents the solution of the system defined
                     by the Hamiltonian $H_{QE}(0,\beta)$ (note that Hamiltonian $H_{QE}$ 
                     depends on $\beta$, hence so is $\rho_t$),
 \end{enumerate}
then the reduced dynamics $\eta_t$ for the model governed by the Hamiltonian~(\ref{ht})
can be obtained using the following simple formula  

\begin{equation}
 \label{r:rho}
   \eta_t = V_t\rho_t\left(\beta_{eff}\right)V_t^{\dagger}, 
   \quad \beta_{eff}:=\beta-\omega/2,
\end{equation}
where the unitary matrix $V_t$ is given by (diag-diagonal)

\begin{equation}
 \label{r:v}
     V_t = \mbox{diag}\left(e^{-i\omega t/2},e^{i\omega t/2}\right). 
\end{equation}
Equality~(\ref{r:rho}) states that having the reduced dynamics $\rho_t$ one may
easily obtain the solution we are interested in simply by introducing the
\emph{effective} parameter $\beta_{eff}$, then replacing $\beta$ by
$\beta_{eff}$ and finally performing the unitary transformation~(\ref{r:v}). 

The interaction term defined in the Eq.~(\ref{ising}) satisfies the requirement of the
theorem above, thus one should restrict the analysis to the Hamiltonian $H_{QE}(\beta) \equiv
H_{QE}(0,\beta)$. The later can be easily rewritten as a $2\times 2 $ block
operator matrix, namely

\begin{equation} 
 \label{total}
   H_{QE}(\beta)
           = 
           \begin{bmatrix}
                H_{\smal{+}}(\beta) & \alpha\idE \\ 
                \alpha\idE &   H_{\smal{-}}(\beta)
            \end{bmatrix},
\end{equation}
where

\begin{equation}
     H_{\smal{\pm}}(\beta):=
     \sum_{n=1}^{N}\left(\omega_{n}^{\smal{\pm}}\sigma_n^z\pm\bar{\beta}\idE\right),
     \quad \bar{\beta}:=\beta/N,
\end{equation}
and $\omega_n^{\smal{\pm}}:=\omega_n\pm g_n$, for each $n\le N$. Henceforward,
the explicit dependence of any quantity of parameter $\beta$ will be omitted until
the solution $\rho_Q(t)$ is obtained.

The evolution operator for the total system generated by the
Hamiltonian~(\ref{total}) reads

\begin{equation}
 \label{evolv}
   \begin{split}
           U_{t} &=\exp\left(-iH_{QE}t\right) \\
                 &=U_{X}\exp\left(-iH_{d}\right)U_{X}^{-1},
   \end{split}
\end{equation}
where $U_{X}$ is given by the Eq.~(\ref{ux}) and $X$ is the solution of the
Riccati Eq.~(\ref{Riccati}), which in the present case takes the form

\begin{equation}
  \label{ariccati}
      \alpha X^2+XH_{+}-H_{-}X-\alpha\idE=0.
\end{equation}
The quantity $H_{d}$ represents a diagonal form of the operator matrix $H_{QE}$.
According to the Eq.~(\ref{xdiag}) it reads

\begin{equation}
\label{hdiag}
H_d
  =
   \begin{bmatrix}
     H_{+}+\alpha X & 0_{E} \\
     0_{E} & H_{-}-\alpha X^{\dg}
   \end{bmatrix}.
\end{equation}
Note that for $\alpha=0$ the solution (possibly not the only one) of the
Eq.~(\ref{ariccati}) is given by $X_0=0_{E}$. It is obvious, since in that case
$H_{QE}$ is already in the diagonal form, \ie, $H_{QE}=H_d=\mbox{diag}(H_{\smal{+}},
H_{\smal{-}})$. In order to obtain the solution for $\alpha\not=0$ more subtle 
investigation, provided in the next subsection, is needed.
   
\subsubsection{Solution of the Riccati equation for the \texorpdfstring{$\alpha\not=0$}{} case} 

First, one can observe that $[H_{\smal{-}},H_{\smal{+}}]=0$. Since the solution $X$
of the Eq.~~(\ref{ariccati}) is a function of the operators $H_{\smal{\pm}}$, then
it also must commute with those operators. The Riccati equation in question can be
simplified to the more compact form, namely

\begin{equation}
 \label{vriccati}
     \alpha X^{2}+2VX-\alpha\idE =0,
     \quad\alpha\not=0,
\end{equation}
where we introduced

\begin{equation}
    V = \sum_{n=1}^{N}\left(g_n\sigma_n^z+\bar{\beta}\idE\right).
\end{equation}
Let us assume that the eigenvalue problem for the operator $V$ can be (easily)
resolved. If this is the case we can write the solution $X$ of the Eq.~(\ref{vriccati})
in a manageable form using the spectral theorem for the Hermitian operators. By doing
so we obtain that $X=f(V)$, where

\begin{equation}
 \label{fl}
    f(\lambda)= \frac{\sqrt{\lambda^2+\alpha^2}-\lambda}{\alpha}, 
    \quad\lambda\in\sigma(V).
\end{equation}
We can also represent the solution $f(V)$ in an equivalent way, \ie

\begin{equation}
  \label{fv}
     f(V) = \sum_{\lambda\in\sigma(V)}f(\lambda)\ket{\lambda}\bra{\lambda}.
\end{equation}
Note that $f=f^{\ast}$, \ie, $f$ is a real function for any value of the
parameters $\alpha$ and $\beta$. Here, by $\sigma(V)$ we denoted the spectrum of
$V$. We want to emphasize that the representation~(\ref{fl}) or~(\ref{fv}) of
the solution $X$ of the Eq.~(\ref{vriccati}) is useful computationally if the eigenvalues
of $V$ can be computed. We show below that this is the case. Note also that $X=X^{\dg}$
iff $V=V^{\dagger}$.

Let $i=i_1i_2...i_N$ be a binary ($i_n=0,1$, for $n\le N$) expansion of an
integer number $i\in[0,2^N-1]$. Clearly, the set of all eigenstates $\ket{i}=
\ket{i_1}\ot\ket{i_2}\ot\ldots\ot\ket{i_N}$ forms an orthonormal basis in
$\mc{H}_{E}$. Moreover, because of the following equality
$\sigma^z\ket{i_n}=(-1)^{i_n}\ket{i_n}$ we have

\begin{equation}
  \label{Ei}
    \begin{split}
                V\ket{i} &=\sum_{n=1}^{N}\left(g_n(-1)^{i_n}+\bar{\beta}\right)\ket{i} \\
                         &\equiv E_i\ket{i}, \quad 0\le i\le 2^N-1.
   \end{split}
\end{equation}
Thus, the eigenvalue problem for $V$ has been solved.
Therefore, the solution of the Riccati Eq.~(\ref{vriccati})
$X$ reads

\begin{equation}
  \label{xriccati}
      X =\sum_{i=0}^{2^N-1}f(E_i)\ket{i}\bra{i}. 
\end{equation}

At this point two remarks should by made. The first one is that the
operator~(\ref{xriccati}) depends on the parameter $\alpha$, \ie, 
$X=X(\alpha)$ and $X(\alpha)\rightarrow X_0$ as $\alpha$ goes to $0$.
To see that this statement holds we rewrite the Eq.~(\ref{fl}) in the form

\begin{equation}
   f(\lambda) = \frac{1}{\sqrt{(\frac{\lambda}{\alpha})^2+1}+\frac{\lambda}{\alpha}},
   \quad \lambda\in\sigma(V).
\end{equation}  
Clearly, $f(\lambda)\rightarrow 0$ as $\alpha\rightarrow 0$, unless $\lambda=0$.
The later means that the solution $X$ of the Riccati Eq.~(\ref{vriccati}) is a
continuous function of the parameter $\alpha$, including the $\alpha=0$ value.
From the physical point of view it means that one can control energy exchange
between the systems in a continuous way. Note that if $\alpha=0$ then the energy
transfer is absent. This scenario can be accomplished by taking limits
$\alpha\rightarrow0$ with the final solution we are about to give. Note also
that since for every $\alpha\not=0$ and each $\lambda\in\sigma(V)$ the
eigenvalues $f(\lambda)$ of the operator $X$ we constructed are positive. 
This solution is a positively defined operator on $\mc{H}_{E}$.  

The second remark is that there exists, at least one more operator function that
satisfies the Eq.~(\ref{vriccati}), namely the one specified by the following formula:
$f_{2}(\lambda)=(-\sqrt{\lambda^2+\alpha^2}-\lambda)/\alpha$ for $\lambda\in\sigma(V)$.
This solution represents negatively defined operator. The choice between the different
solution of the Eq.~(\ref{vriccati}) is \emph{not} arbitrary and it has significant
influence on the further analysis. Indeed, if one decides to choose the second solution:
 $f_2(\lambda)$ then one may meet a serious difficulties during the examination of the 
dephasing phenomena, since $f_2({\lambda})\to -\infty$ as $\alpha\to 0$.  
    
At the end of this subsection let us note that one can resolve the eigenvalue
problem for $H_{\smal{\pm}}$ as easily as for the operator $V$. Indeed, we may
readily verify that

\begin{equation}
  \label{Epm}
    \begin{split}
         H_{\smal{\pm}}\ket{i}&=\sum_{n=1}^{N}\left(\omega_n^{\smal{\pm}}(-1)^{i_n}
                                \pm\bar{\beta}\right)\ket{i} \\
                              &\equiv E_i^{\smal{\pm}}\ket{i}, \quad 0\le i\le 2^N-1.  
   \end{split}
\end{equation}
Observe also that $E_i^{\smal{+}}-E_i^{\smal{-}}=2E_i$, for $0\le i\le 2^N-1$,
as one may expected. In the $\ket{i}$ basis the density matrix $\rho_E$ can be
expanded as

\begin{equation}
 \label{rhoE1}
        \rho_{E} = \sum_{i=0}^{2^N-1}\left(e^{-\Omega_i/kT}/Z\right)\ket{i}\bra{i},
\end{equation}
where $\Omega_i=\sum_n\omega_n(-1)^{i_n}$ are the eigenvalues of $H_{E}$ and
$Z=\sum_i\exp(-\Omega_i/kT)$.

\subsubsection{Total system evolution}

We are now in a position to give the explicit and manageable form of the
evolution operator of the total system $Q+E$. First, let us note that the
similarity operator matrix $U_{X}$ can by written as

\begin{equation}
 \label{fux}
    U_{X}=\sum_{i=0}^{2^N-1}U_i\ot\ket{i}\bra{i},
\end{equation}
where matrices $U_i$ are given by

\begin{equation}
 \label{similarU}
   U_i =
       \begin{pmatrix}
         1 & -f(E_i) \\
         f(E_i) & 1
       \end{pmatrix}.
\end{equation}
This follows immediately from the Eqs.~(\ref{ux}) and (\ref{xriccati}). We
also used the resolution of the identity $\idE$ in the $\ket{i}$ basis. From
the expansion~(\ref{fux}) one can readily see that the inversed operator
$(U_{X})^{-1}$ takes the form

\begin{equation}
(U_{X})^{-1}=\sum_{i=0}^{2^N-1}(U_i)^{-1}\ot\ket{i}\bra{i},
\end{equation}
where $(U_i)^{-1}$ stands for the inverse of the matrix $U_i$. Since for every
integer $i$ we have $\mbox{det}(U_i)=1+f^2(E_i)\not=0$, thus the inverse exists.
It is also easy to see that $(U_i)^{-1}=\det(U_i)U_i^{\dg}$ and $U_iU_i^{\dg}=1/
\mbox{det}(U_i)$. Because of the last equality, the operator~(\ref{fux}) can become
unitary by a simple rescaling procedure, \ie, $U_i\rightarrow\sqrt{
\mbox{det}U_i}U_i$. Furthermore, according to the Eq.~(\ref{evolv}) the diagonal form
$U_t^d:=\exp(-iH_dt)$ reads

\begin{equation}
   U^d_t=\sum_{i=0}^{2^N-1}U_i^d(t)\ot\ket{i}\bra{i}, 
\end{equation}
where abbreviations

\begin{equation}
   U_i^d(t)= \mbox{diag}(e^{-i(E_i^{+}+\alpha f(E_i))t},e^{-i(E_i^{-}-\alpha f(E_i))t}),
\end{equation}
were introduced. 

Finally, by combining all the results together we obtain the evolution operator
$U_t$ of the total system:

\begin{equation}
 \label{ut}
    U_t = \sum_{i=0}^{2^N-1}U_i(t)\ot\ket{i}\bra{i},
\end{equation}
where

\begin{equation}
 \label{iunitary}
   \begin{split}
        U_i(t) &= U_i U_i^d(t)U_i^{\dagger} \\
               &= \frac{1}{1+f_i^2}
 \begin{pmatrix}
  e_i^{\smal{+}}(t)+e_i^{\smal{-}}(t)f_i^2 & f_i(e_i^{\smal{+}}(t)-e_i^{\smal{-}}(t)) \\
  f_i(e_i^{\smal{+}}(t)-e_i^{\smal{-}}(t)) & e_i^{\smal{+}}(t)f_i^2+e_i^{\smal{-}}(t) 
 \end{pmatrix}.
   \end{split}
\end{equation}
and $e_i^{\smal{\pm}}(t):=\exp(-i(E_i^{\smal{\pm}}\pm\alpha f_i)t)$ with
$f_i\equiv f(E_i)$. Let us keep in mind that $U_iU_i^{\dg}=\idQ$ for each
integer $i$.

\subsubsection{Reduced dynamics}
  \label{RD}

We begin with the assumption that $\rho_Q$ and $\rho_E$ are arbitrary density
operators for the open system and its environment, respectively. Furthermore, we
consider only factorable initial states $\rho_{QE}$ of the closed system $Q+E$,
\ie, $\rho_{QE}=\rho_{Q}\ot\rho_{E}$. Using this assumptions and the
explicit form of the evolution operator $U_t$ given in the Eq.~(\ref{ut}), we have
that the density operator $\rho_{QE}(t)=U_t\rho_{QE}U_t^{\dagger}$ at any given
time reads

\begin{equation}
  \label{rhoqet}
\rho_{QE}(t)=
\sum_{i,j=0}^{2^N-1}U_i(t)\rho_QU_j(t)^{\dg}\ot\ket{i}\bra{i}\rho_E\ket{j}\bra{j}.
\end{equation}   
Tracing out the last equation over the environment degrees of freedom, we obtain 
the reduced dynamics (note $\rho_{Q}(t)=\mbox{Tr}_{E}(\rho_{QE}(t))$)

\begin{equation}
\label{rhoqt}
 \rho_{Q}(t)=
 \sum_{i=0}^{2^N-1}\rho_iU_i(t)\rho_QU_i(t)^{\dg},
\end{equation}
where $\rho_i\equiv\bra{i}\rho_E\ket{i}$. By introducing matrices
$K_{ij}(t)$ such that $K_{ij}(t):=\delta_{ij}\sqrt{\rho_i}U_i(t)$, the 
equation above can be written in a more familiar form, that is
$\rho_{Q}(t)=\sum_{ij}K_{ij}(t)\rho_QK_{ij}(t)^{\dg}$. Moreover, it is easily to
verify that $\sum_{ij}K_{ij}(t)K_{ij}(t)^{\dg}=\idQ$. Thus, the Eq.~(\ref{rhoqt})
can be thought of as operator sum representation of the state $\rho_Q(t)$. The
matrices $K_{ij}(t)$ are the Kraus operators.  
The result~(\ref{rhoqt}) is general, \ie, it holds for the arbitrary states $\rho_Q$ and
$\rho_E$. Nevertheless, we made assumption about the form of the initial
state $\rho_{QE}$. This assumption, however, is not essential and does not
lead to the limitation of the analysis. The evolution operator~(\ref{ut}) was written
in a highly manageable form, and by that we mean it can be applied to \emph{any} given
initial state. Of course, if initial correlations are present, then the Eqs.~(\ref{rhoqet}),
and ~(\ref{rhoqt}) no longer hold, but the reduced dynamics $\rho_Q(t)$ can still
easily be obtained. However, we will not focus on this issue in the current paper.

So far, we omitted the explicit dependency on $\beta$ in the introduced quantities.
Note, however, that if $U_i(t)=U_i(t,\beta)$,  then $\rho_{Q}(t)=\rho_{Q}(t,\beta)$ and
in view of the Eqs.~(\ref{r:rho}),~(\ref{rhoqt}),~(\ref{r:v}) and comments following them,
we have

\begin{equation}
 \label{eta}
 \eta_t =  \sum_{i,j=0}^{2^N-1}M_{ij}(t)\eta_Q M_{ij}(t)^{\dg},
\end{equation}
where $\eta_Q$ ($\eta_Q=\rho_Q$) is an arbitrary initial state. The Kraus matrices $M_{ij}(t)$
are given by
\begin{equation}
 \label{etaKraus}
 M_{ij}(t) = \delta_{ij}\sqrt{\rho_i}V_tU_i(t,\beta_{eff}).
\end{equation}

The last two equations are the main result of the current paper.
It is worth mentioning that the model with a similar properties was studied
in~\cite{chaotic}, where the Authors obtained the operator sum representation, yet
the Kraus operator introduced therein involved the time chronological operator.

From the Eqs.~(\ref{eta}) and~(\ref{etaKraus}) one can learn that the time evolution
of the qubit interacting with the rotating magnetic field~(\ref{field}) and in
contact with the environment~(\ref{hr})~-~(\ref{ising}) can be described using
two quantum channels, \ie, $\rho_Q(t)=T_t^2\circ T_t^1(\rho_Q)$. The first channel,
that is $T_t^1$ is given by

\begin{equation}
 \label{channel1}
  T_t^1:=\sum_{i=0}^{2^N-1}M_{i}(t)(\cdot)M_{i}(t)^{\dagger}.
\end{equation}
The second one is defined by the simple unitary operation, namely $T_t^2=V_t(\cdot)
V_t^{\dagger}$. The matrices $M_i$ are specified by

\begin{equation}
 \label{HKraus}
  M_i(t)=\sqrt{\rho_i}\exp(-iH_it),  
\end{equation}
where the Hamiltonian $H_i$ is the generator of the unitary evolution
$U_i(t,\beta_{eff})$ (see~(\ref{iunitary})). It is easy to compute that

\begin{equation}
 \label{Hi}
 H_i = 
 \begin{pmatrix}
 E_i-\omega/2 & \alpha \\
  \alpha & -E_i+\omega/2
 \end{pmatrix}, 
\end{equation}
where $E_i$ is the eigenvalues of the operator $V$ and it was defined in the
Eq.~(\ref{Ei}). The channel $T_t^2$ may by thought of as the convex
combination of the unitary channels (note $T_t(\idQ)=\idQ$). This type of 
channels are known as the \emph{random unitary channels}~\cite{unital} and 
they often appears in the study of the pure decoherence.

Surprisingly, the matrix above and therefore the Kraus matrices~(\ref{HKraus})
does \emph{not} depend on the function $f(E_i)$. However, to compute matrices
$\exp(-iH_it)$ that specified the Kraus matrices~(\ref{HKraus}) we need to
diagonalize the Hamiltonians $H_i$. The function $f(E_i)$ is hidden in the
similarity matrix $U_i$ given by the Eq.~(\ref{similarU}). Observe that the
mentioned diagonalization procedure is based on the Riccati equation (matrices
$U_i$ are composed with the eigenvalues of the solution of the Riccati equation).
This approach differs from the standard method based on the characteristic equation.
This is a new kind of the diagonalization procedure, so called Riccati
diagonalization~\cite{Rdiag}.

\section{Special cases}

Now, we look into some common cases that may arise during the examination of
the model~(\ref{ht}). We will show that in these particular situations the
result~(\ref{eta}) can be simplified to the well-know expressions.

\subsection{No coupling with the bath}
 
In the case when the qubit under consideration has no coupling with its
environment, \ie, $g_n=0$ for all $n\le N$ we have
 \begin{enumerate}
   \item[$a)$]  $E_i=\beta_{eff}$, hence
   \item[$b)$]  $f_i\equiv f=(|z|-\beta_{eff})/\alpha$, where
   \item[$c)$]  $z:=\alpha+i\beta_{eff}$.
 \end{enumerate}
Moreover, $E_i^{\smal{\pm}}=\Omega_i\pm\beta$, where
$\Omega_i\equiv\sum_n\omega_n(-1)^{i_n}$. Using this simplification we obtain
that $U_i(t,\beta_{eff}) = e^{-i \Omega_it} \bar{U}(t)$, where
\begin{equation}\label{ubar}
  \begin{split}
  \bar{U}(t) &= 
  \begin{pmatrix}
  e^{-i|z|t}+f^2e^{i|z|t} & -2if\sin(|z|t) \\
  -2if\sin(|z|t) & f^2e^{-i|z|t}+e^{i|z|t}
  \end{pmatrix} \\
  &=
  I_2\cos(|z|t)-i
  \begin{pmatrix}
  \frac{1-f^2}{1+f^2} & \frac{2f}{1+f^2}  \\
  \frac{2f}{1+f^2} & -\frac{(1-f^2)}{1+f^2}
  \end{pmatrix}
  \sin(|z|t)    \\
   &= I_2\cos(|z|t)-i\bs{\sigma}\cdot\hat{\bs{\Omega}}\sin(|z|t) \\
   &= \exp(-i|z|\bs{\sigma}\cdot\bs{\hat{\Omega}}t),
  \end{split}
\end{equation}
with $\hat{\bs{\Omega}}:=\bs{\Omega}/\|\bs{\Omega}\|$ and
$\bs{\Omega}:=(2f,0,1-f^2)$. As a result, 
$M_{ij}(t)=\delta_{ij}\sqrt{\rho_i}V_t\bar{U}(t)$, where we dropped phase
factor $e^{-i \Omega_it}$. Therefore $\eta_t =U(t)\eta U(t)^{\dagger}$ (as one
may anticipated), where

\begin{equation}
  \label{floquet}
  U(t) = \exp(-i\omega t\sigma^z/2)\exp(-i|z|\bs{\sigma}\cdot\bs{\hat{\Omega}}t).
\end{equation}
This is the well-known formula for the (unitary) evolution operator of the system
identified by the Hamiltonian $H_{Q}(t,\beta)\equiv H_{Q}(t)$. It is the solution
of the differential equation $i\partial_tU(t)=H_{Q}(t)U(t)$. Note, that the
result~(\ref{floquet}) corresponds to the Floquet decomposition of the unitary
operator generated by the periodic Hamiltonian. Note also that the evolution
operator~(\ref{floquet}) can by written in a more readable form, especially
useful for studying the adiabatic approximation (see Sec.~(\ref{sec:adiab})), \ie

\begin{equation}
 \label{useful}
 U(t)= V_t \exp(-iHt),
\end{equation}
where the matrix $V_t$ is given by the Eq.~(\ref{r:v}) and $H$ has the form

\begin{equation}
 H = 
 \begin{pmatrix}
 \beta_{eff} & \alpha \\
 \alpha & - \beta_{eff}
 \end{pmatrix}.
\end{equation}
One can easily see that this matrix can be obtained directly from the Eq.~(\ref{Hi})
if one set $g_n=0$ for $n\le N$.


\subsection{Dephasing}
  
If the constant $\alpha$ equals zero, then the Hamiltonian $H_{QE}(t,\beta)$ becomes
time-independent. Moreover, in this circumstances $[H_Q,H_{int}]=0$ hence the open
system does not exchange energy with its environment, \ie, the pure decoherence
occurs. It poses no problems to check that $U_i(t,\beta_{eff})=e^{-i\Omega_it}V_t^{\dg}
\exp(-iE_i\sigma^zt)$. Therefore, the Kraus operators take the form
$M_{ij}(t)=\delta_{ij}\sqrt{\rho_i}\exp(-iE_i\sigma^zt)$. Just like before the phase
factor $e^{-i \Omega_it}$ was drooped. In agreement with~(\ref{eta}) the form above
of the Kraus matrices leads to the following manifestation of the density matrix
$\eta_t$

\begin{equation}
 \eta_{11}(t)=\eta_{11},\quad 
 \eta_{12}(t)=\sum_{i=0}^{2^N-1}\rho_i e^{-i2E_it}\eta_{12} 
\end{equation}
Naturally, $\eta_{22}(t)=1-\eta_{11}(t)$ and $\eta_{21}(t)=\eta_{21}(t)^{\ast}$.
Note that the coherence $C(t):=|\eta_{12}(t)|$ does not decay exponentially (or
anyhow for finite $N$) in the long time regime, but manifests oscillating
behavior.

From the considerations above one can easily learn that there is a substantial
difference between the situation when a transfer of the energy from the qubit to its
bath is not present and the case when the qubit is not coupled to the environment. 
Let us also observe that in the model we study it is not possible to construct a
situation when the energy exchange in not present $(\alpha=0)$ and yet the magnetic
field is rotating. This follows from the fact that if $\alpha=0$ then the
Hamiltonian $H_{QE}(t,\beta)$ does not depend on $\omega$.

\subsection{Low temperature regime}
  
From the mathematical point of view the zero (low) temperature limit is the most
subtle special case. This comes from the fact that most results obtained by
using the known approximation methods in the open quantum systems theory, \eg,
the Markovian or the singular coupling limit, cannot by extrapolated to this
regime. Luckily, the exactly solvable models are unhampered by this
difficulties.

To derive the exact reduced dynamics in the low temperature domain we begin with
rewriting the density operator $\rho_{E}$ in a more suitable form in contrary with
the one given by Eq.~(\ref{rhoE1}). According to the result obtained in~\cite{spinStar1} 
one may write

\begin{equation}
  \label{rhoE2}
    \rho_E= \bigotimes_{n=1}^{N}\frac{1}{2}\left(\idQ+\beta_n\sigma^z\right),
\end{equation}  
where $\beta_n=\tanh(-\omega_n/kT)$. From equality~(\ref{rhoE2}) one may readily see
when $T\to 0$, then

\begin{equation}
    \rho_i=\frac{1}{2^N}\prod_{n=1}^{N}\left(1+\beta_n(-1)^{i_n}\right)\to\delta_{i0},
\end{equation}
since $\beta_n\to 1$ as $T\to 0$. All this means that in the low temperature
domain the heat bath is its ground state, \ie, $\rho_E=\ket{0_E}\bra{0_E}$.
As an immediate result we obtain the following form of the reduced dynamics~(\ref{eta})

\begin{equation}
   \lim_{T\to0}\eta_t =  W_t\eta W_t^{\dagger},
\end{equation}
where $W_t:=V_tU_{0}(t,\beta_{eff})$. Interestingly, the evolution of the open
system in question is unitary, which is rather unusual. This is a direct consequence 
of the fact that for every integer $i$ the operator $U_i(t)$ specified
by the Eq.~(\ref{iunitary}) is unitary. This means that in the low temperature regime 
the dissipative properties of the heat bath~(\ref{hr})~-~(\ref{ising}) are frozen.

\subsection{Equal Couplings and Frequencies}

Henceforward and without essential loss of generality we assume that
$\omega_n=\Omega$ and $g_n=g$ for $n\le N$, \ie, couplings constants and the
frequencies of the spins of the bath are equal. In this situation
$E_i=g(N-2k)+\beta$, where $k$ is the \emph{Hamming weight} of the integer number
$i$ (\ie, a number of nonzero element in a binary expansion of $i$). Since there
are $\tbinom{N}{k}$ integer number $i\in [0,2^N-1]$ with the same Hamming weight
$k$, the channel~(\ref{channel1}) takes the form

\begin{equation}
 T_t^{1} = \sum_{k=0}^N\dbinom{N}{k}M_k(t)(\cdot)M_k(t)^{\dagger},
\end{equation}
with $M_k(t)$ given by Eq.~(\ref{HKraus}). Note, in this case $\rho_k$ (see
Eq.~(\ref{rhoE2})) simplified to the form

\begin{equation}
 \rho_k = \frac{1}{2^N}(1+\delta)^{N-k}(1-\delta)^k,
\end{equation} 
where $\delta:=\tanh(-\Omega/kT)$.
 

\section{Application: adiabatic approximation}
  \label{sec:adiab}

The adiabatic theorem~\cite{MSAdiab,TongAdiab} for the \emph{closed} quantum
system specified by the Hamiltonian $H(t)$ in its basics variant (\ie, discrete
and no degenerate spectrum $\sigma(H(t))$) states that if $H(t)$ varies slowly
(for rigorous meaning of that see \eg,~\cite{Avron} and references therein) and if
the system is initially prepared in one of the eigenstates of $H(0)$, say
$\ket{\psi_n(0)}$, then at any given time $t$ the probability
$|\iner{\psi_n(t)}{\psi(t)}|^2$ of finding it in the eigenstate
$\ket{\psi_n(t)}$ (an eigenvector of $H(t)$) is equal to $1$. Therefore, one may
easily introduce the quantity $F(t)$ that measures the validity of the adiabatic
approximation, namely

\begin{equation}
  \label{adiab}
    F(t)=\mbox{Tr}(\rho(t)\rho_{\psi}(t)),
\end{equation}
where $\rho_{\psi}(t)=\ket{\psi_t}\bra{\psi_t}$ and
$\rho(t)=U_t\rho(0)U_t^{\dagger}$ with $\rho(0)=\ket{\psi_0}\bra{\psi_0}$. Here,
$\ket{\psi_t}$ is the eigenvector of $H(t)$ for $t\ge0$ and $U_t$ represents the
(unitary) evolution operator for the closed system. If the system is evolving
adiabatically then $F(t)= 1$. Note that definition of the quantity $F(t)$
coincides with the standard definition of the quantum fidelity
$F(\rho(t),\rho_{\psi}(t))$ between \emph{pure} states $\rho(t)$ and
$\rho_{\psi}(t)$~\cite{fidelity}. Since this is a two dimensional case it also
coincides with the definition of the super fidelity~\cite{JarekSF,sub}.

Observe that the Eq.~(\ref{adiab}) is very useful since it can be easily applied
both to the closed and to the open quantum systems. In the last case we
need to replace the equality $\rho(t)=U_t\rho(0)U_t^{\dagger}$ representing the
evolution of the closed system by the channel describing evolution in time of
the open system. More precisely, $\rho(t)=T_t(\rho(0))$, where $T_t$ is TP-CP
quantum operation. In the current section we use the Eq.~(\ref{adiab}) in order
to investigate behavior of the model~(\ref{ht}) in the adiabatic regime. To
make the paper self-contained we briefly discuss the adiabatic approximation for
the system~(\ref{spin}) first.  

\subsection{The closed spin system case}
 
For the sake of simplicity we put $\beta=\beta_0\cos\varphi$ and
$\alpha=\beta_0\sin\varphi$ for certain $\varphi$ and $\beta_0$. This
assumptions lead to the following form of the Hamiltonian~(\ref{spin2})

\begin{equation}
 \label{spin3}
H_{Q}(t) = \beta_0
  \begin{pmatrix}
   \cos\varphi & e^{-i\omega t}\sin\varphi \\
    e^{i\omega t}\sin\varphi &  -\cos\varphi
   \end{pmatrix}.
\end{equation}

It can be easily verified that the matrix above has the eigenvalues
$E_{\smal{\pm}}=\pm\beta_0$. The corresponding eigenvectors
$\ket{\psi_t^{\smal{\pm}}}$ are given by

\begin{equation}
 \ket{\psi_t^{+}}
=
\begin{pmatrix}
  \cos(\varphi/2)  \\
  e^{i\omega t}\sin(\varphi/2) 
 \end{pmatrix},
\quad 
 \ket{\psi_t^{-}}
=
\begin{pmatrix}
  \sin(\varphi/2)  \\
  -e^{i\omega t}\cos(\varphi/2) 
 \end{pmatrix}.
\end{equation}
The density matrix $\rho_{+}(t)=\ket{\psi_t^{+}}\bra{\psi_t^{+}}$ takes the form

\begin{equation}
 \rho_{\smal{+}}(t) = 
 \begin{pmatrix}
 \cos^2(\varphi/2) & \frac{1}{2}e^{-i\omega t}\sin\varphi \\
 \frac{1}{2}e^{i\omega t}\sin\varphi & \sin^2(\varphi/2)
 \end{pmatrix}.
\end{equation}
Observe that $\rho_{+}(t)=V_t\rho_{+}(0)V_t^{\dagger}$, where $V_t$ is specified
in the Eq.~(\ref{r:v}). According to the Eq.~(\ref{useful}) the density matrix $\rho(t)$
at any given time $t$ reads $\rho(t)=V_t\bar{U}(t)\rho_{+}(0)(V_t\bar{U}(t))^{\dagger}$,
thus the fidelity~(\ref{adiab}) takes the form

\begin{equation}
 \label{adiab2}
  \begin{split}
F(t) &= \mbox{Tr}(\bar{U}(t)\rho_{\smal{+}}(0)\bar{U}(t)^{\dagger}\rho_{\smal{+}}(0)) \\
     &= \|\bar{U}(t)\rho_{\smal{+}}(0)\|_{F}^2,
  \end{split}
\end{equation}
where $\|\cdot\|_{F}$ is the Frobenius (Euclidean) norm, \ie, $\|A\|_{F}^2:=\mbox{Tr}(AA^{\dagger})$.

It is well-known that for the system governed by the Hamiltonian~(\ref{spin3})
condition that guarantees the adiabatic evolution is $\beta_0\gg\omega$. This
statement is very intuitive, it means that the magnetic field rotates slowly, in
comparison with the phase of the state vector (for more detail discussion
see \eg,~\cite{griff}). We can also see this from the Eq.~(\ref{adiab2}). Indeed, if one
introduces the \emph{adiabatic parameter} $x:=\omega/2\beta_0$, then in
agreement with~(\ref{useful}) and~(\ref{adiab2}) we have

\begin{equation}
 \label{xadiab}
 F(t) =1-\frac{x^2}{1+x^2}\sin^2(\Omega(x)t), 
\end{equation}
where $\Omega(x):=\beta_0\sqrt{1+x^2}$. Without loss of generality we set
$\varphi=\pi/2$ in the equation above. The parameter $x$ measures how slowly the
magnetic field rotates in $\beta_0$ units, thus in the adiabatic limits (\ie,
$x\to 0$) the second term in the Eq.~(\ref{xadiab}) can be neglected and therefore
$F(t)\simeq 1$.
 
\subsection{The open system case}
 
By analogy to the closed spin case discussed above one may write the
fidelity~(\ref{adiab}) for the open system~(\ref{ht}) as 

 \begin{equation}
   \label{Fopen}
     \begin{split}
        F(t) &= \mbox{Tr}(T_t^2\circ T_t^1(\rho_{\smal{+}}(0))\rho_{\psi}(t)) \\
             &=\sum_{k=0}^{N}\dbinom{N}{k}\rho_k F_k(t), 
   \end{split}
 \end{equation} 
where $F_k(t):=\|U_k(t)\rho_{\smal{+}}(0)\|_{F}^2 $. The channels $T_t^1$ and
$T_t^2$ are specified in the Eq.~(\ref{channel1}) and comments bellow, respectively.
The unitary matrix $U_k$ takes the form $U_k=\exp(-iH_kt)$, with $H_k$ given by
the Eq.~(\ref{Hi}). If we put $\varphi=\pi/2$ as  before, then 

\begin{equation}
  \label{xkadiab}
     F_k(t) = 1- \frac{x_k^2}{1+x_k^2}\sin^2(\Omega(x_k) t),
\end{equation}
where $x_k:=G(N-2k)/\beta_0-x$. Using this and Eq.~(\ref{Fopen}) we obtain that
$F(t)=1-R(t)$, where $R(t)$ reads

\begin{equation}
R(t) = \sum_{k=0}^{N}\dbinom{N}{k}\frac{x_k^2}{1+x_k^2}\rho_k\sin^2(\Omega(x_k) t).
\end{equation}
From this equation one can readily see that in the adiabatic domain ($x\to 0$)
we have $R(t)\not=0$ and thus $F(t)<1$. Therefore, the standard condition that
leads to the adiabatic behavior of the closed system~(\ref{spin}) does \emph{not}
guarantee that in the case of the open system model~(\ref{ht}) it holds true as well.
However, if one additionally assumes that coupling with the
environment is weak, in comparison with the energy split between the states
$\ket{0}$ and $\ket{1}$ of the qubit in question, that is to say if
$G/\beta_0\ll 1$, then $x_k\ll 1$ (for finite $N$) and $R(t)\simeq 0$,
thus $F(t)\simeq 1$.
  
\section{Summary}
 \label{sec:summary}

In this paper we investigated a qubit in contact with the spin environment and interacting
with a rotation magnetic field. The considered model was constructed under a set of assumptions
which allow for its exact treatment. We hope that despite the mathematical character the results
of the paper may serve as starting point for further investigations. The exact models can not 
only provide reasonable approximate description of real systems, as it is the case for the pure
dephasing, but are often used as a basis and inspiration for various improvements~\cite{X}.
Although the paper is mainly focused on mathematical aspects it also includes the example of a
natural application of the model in question. This example relates our model to important 
problems of physics such as \eg~the problem of the adiabatic approximation for the open quantum
systems~\cite{AA} or the adiabatic quantum computation~\cite{AQQ}.

We provided
the exact reduced dynamics for the system mentioned above. In contrast to the standard methods
available in the theory of the open quantum systems our approach was based on the
block operator matrices theory. In particular, we resolved the algebraic 
Riccati equation associated with the Hamiltonian defining the model under consideration.
We wish to stress out that the method we used in the current paper, although applied
to the particular model, is general. Nevertheless, its usefulness relies on the ability of
solving the Riccati equation. At the present time it is a very difficult task, even for the
simple systems.

At this point one can ask: how relevant is it to assume a spin environment, instead
of a bosonic one, to solve the problem exactly? Is it possible to do so for bosonic
bath? One may pose a more general question. What is a connection between the possibility
of obtaining the exact reduced dynamics of the qubit in question and the Hamiltonian 
specifying the bath? This problem was already addressed in~\cite{gardas}. The results of
this paper as well as the analysis we carried out suggest that this problem is at
least as difficult as resolving the Riccati equation associated with the total Hamiltonian.
 
Furthermore, we studied the adiabatic approximation for the model in question.
It was shown that the standard condition that guarantees the adiabatic evolution
in the the case of the closed systems is not valid for the open system generalization.
This is not an unexpected result. It is interesting, however, that the aforementioned
condition does ensures the adiabatic behavior of the open system under consideration
in the weak coupling limit.

\begin{acknowledgments}
The author would like to thank the anonymous referees for their comments. The author 
also thank Jerzy Dajka and Jaros\l{}aw Adam Miszczak for helpful suggestions. This 
work was supported by the Polish Ministry of Science and Higher Education under the
grant number N N$519$ $442339$.
\end{acknowledgments}
   
\appendix
\section{\label{sec:app}Proof of the Eq.~(\ref{r:rho})}

In order to prove the equation~(\ref{r:rho}) let us note that the Hamiltonian
~(\ref{ht}) satisfies the following condition

\begin{equation}
  \label{r:cond}
    H(t,\beta) = e^{iKt}H(\beta)e^{-iKt}, 
\end{equation}
where $K= -\tfrac{\omega}{2}\sigma_3\otimes\idE$. This can be easily proven
using the Baker-Campbell-Hausdorff formula~\cite{galindo}. As was shown, 
in~\cite{thaller} every quantum system with Hamiltonian $H(t,\beta)$ satisfying
~(\ref{r:cond}) for some Hermitian operator $K$  evolves  
\begin{equation}
 \label{r:u}
 U_t(\beta) = e^{iKt}e^{-iH_{eff}(\beta)t},\quad H_{eff}(\beta): = H(\beta)+K. 
\end{equation} 
Note that in general $[H(\beta),K]\not=0$ and therefore $[H_{eff}(\beta),K]\not=0$.
In our case, from equation~(\ref{ht}) we learn that $H(\beta)=
\left(\beta\sigma_3+\alpha\sigma_1\right)\otimes\idE$, thus

\begin{equation}
  \label{r:efektywny}
    \begin{split}
         H_{eff}(\beta) &= \left(\beta\sigma_3+\alpha\sigma_1\right)\otimes\idE
                        -\frac{\omega}{2}\sigma_3\otimes\idE \\ 
                        &= \left(\left(\beta-\frac{\omega}{2}\right)\sigma_3
                        +\alpha\sigma_1\right)\otimes\idE \\ 
                        &= H(\beta-\frac{\omega}{2}). 
    \end{split}
\end{equation}
From equations~(\ref{r:u}) and~(\ref{r:efektywny}) we have

\begin{equation}
 \label{r:evolv}
 U_t(\beta)= e^{iKt}U_t(\beta-\tfrac{\omega}{2}), 
\end{equation}
where $U_t(\beta)$ is the evolution operator generated by $H(t,\beta)$.
Let $\hat{\rho}_t(\beta)$ and $\hat{\eta}_t$ be a density operator for the closed system
$Q+E$ associated with the Hamiltonian $H(\beta)$ and $H(t,\beta)$ respectively in some arbitrary
time $t$. Let us also assume that $\hat{\rho}_0(\beta)=\hat{\eta}_0\equiv \hat{\rho}$. 
Using equation~(\ref{r:evolv}) one can easily see that  
  \begin{eqnarray}
    \label{hat} 
   \hat{\eta}_t & = & U_t(\beta)\hat{\rho}U_t^{\dagger}(\beta) \\ \nonumber
                & = & e^{iKt}U_t(\beta-\tfrac{\omega}{2})\hat{\rho}
                U_t^{\dagger}(\beta-\tfrac{\omega}{2})e^{-iKt} \\ \nonumber
                & = & \hat{V}_t\hat{\rho}_t(\beta-\tfrac{\omega}{2})\hat{V}_t^{\dagger},    
  \end{eqnarray}
where we introduced $\hat{V}_t=e^{iKt}$.
To end the proof we will show that if $\hat{A}_1$, $\hat{A}_2\in B(\mathcal{H}\oplus\mathcal{H})$ 
are a $2\times2$ block operator matrix of the form $\hat{A}_i=A_i\otimes\idE$, $(i=1,2)$ and
$\hat{B}=[\hat{B}_{ij}]\in B(\mathcal{H}\oplus\mathcal{H})$ then
\begin{equation}
 \label{tr}
 \mbox{Tr}_{E}(\hat{A}_1\hat{B}\hat{A}_2)= A_1\mbox{Tr}_{E}(\hat{B})A_2.  
\end{equation}
Equation~(\ref{tr}) follows from the linearity of the trace $\mbox{Tr}(\cdot)$ operation and the
definition~(\ref{partial}) of partial trace. Note that $\hat{V}_t=V_t\otimes\idE$,
where $V_t$ is given by the equation~(\ref{r:v}), thus taking partial trace of the equation
~(\ref{hat}) and using~(\ref{tr}) we obtain~(\ref{r:rho}) with $V_t$ given by~(\ref{r:v}).


\begin{thebibliography}{38}%
\makeatletter
\providecommand \@ifxundefined [1]{%
 \@ifx{#1\undefined}
}%
\providecommand \@ifnum [1]{%
 \ifnum #1\expandafter \@firstoftwo
 \else \expandafter \@secondoftwo
 \fi
}%
\providecommand \@ifx [1]{%
 \ifx #1\expandafter \@firstoftwo
 \else \expandafter \@secondoftwo
 \fi
}%
\providecommand \natexlab [1]{#1}%
\providecommand \enquote  [1]{``#1''}%
\providecommand \bibnamefont  [1]{#1}%
\providecommand \bibfnamefont [1]{#1}%
\providecommand \citenamefont [1]{#1}%
\providecommand \href@noop [0]{\@secondoftwo}%
\providecommand \href [0]{\begingroup \@sanitize@url \@href}%
\providecommand \@href[1]{\@@startlink{#1}\@@href}%
\providecommand \@@href[1]{\endgroup#1\@@endlink}%
\providecommand \@sanitize@url [0]{\catcode `\\12\catcode `\$12\catcode
  `\&12\catcode `\#12\catcode `\^12\catcode `\_12\catcode `\%12\relax}%
\providecommand \@@startlink[1]{}%
\providecommand \@@endlink[0]{}%
\providecommand \url  [0]{\begingroup\@sanitize@url \@url }%
\providecommand \@url [1]{\endgroup\@href {#1}{\urlprefix }}%
\providecommand \urlprefix  [0]{URL }%
\providecommand \Eprint [0]{\href }%
\providecommand \doibase [0]{http://dx.doi.org/}%
\providecommand \selectlanguage [0]{\@gobble}%
\providecommand \bibinfo  [0]{\@secondoftwo}%
\providecommand \bibfield  [0]{\@secondoftwo}%
\providecommand \translation [1]{[#1]}%
\providecommand \BibitemOpen [0]{}%
\providecommand \bibitemStop [0]{}%
\providecommand \bibitemNoStop [0]{.\EOS\space}%
\providecommand \EOS [0]{\spacefactor3000\relax}%
\providecommand \BibitemShut  [1]{\csname bibitem#1\endcsname}%
\let\auto@bib@innerbib\@empty
\bibitem [{\citenamefont {Galindo}\ and\ \citenamefont
  {Pascual}(1990{\natexlab{a}})}]{galindo2}%
  \BibitemOpen
  \bibfield  {author} {\bibinfo {author} {\bibfnamefont {A.}~\bibnamefont
  {Galindo}}\ and\ \bibinfo {author} {\bibfnamefont {P.}~\bibnamefont
  {Pascual}},\ }\href@noop {} {\emph {\bibinfo {title} {Quantum Mechanics vol
  2}}}\ (\bibinfo  {publisher} {Springer-Verlag, Berlin, Germany},\ \bibinfo
  {year} {1990})\BibitemShut {NoStop}%
\bibitem [{\citenamefont {Traller}(2005)}]{thaller}%
  \BibitemOpen
  \bibfield  {author} {\bibinfo {author} {\bibfnamefont {B.}~\bibnamefont
  {Traller}},\ }\href@noop {} {\emph {\bibinfo {title} {Advanced visual quantum
  mechanics}}}\ (\bibinfo  {publisher} {Springer Science+Business Media,
  Inc.},\ \bibinfo {year} {2005})\BibitemShut {NoStop}%
\bibitem [{\citenamefont {{Breuer}}\ and\ \citenamefont
  {Petruccione}(2002)}]{breuer}%
  \BibitemOpen
  \bibfield  {author} {\bibinfo {author} {\bibfnamefont {H.~P.}\ \bibnamefont
  {{Breuer}}}\ and\ \bibinfo {author} {\bibfnamefont {F.}~\bibnamefont
  {Petruccione}},\ }\href@noop {} {\emph {\bibinfo {title} {The Theory of Open
  Quantum Systems}}}\ (\bibinfo  {publisher} {Oxford University Press, USA},\
  \bibinfo {year} {2002})\BibitemShut {NoStop}%
\bibitem [{\citenamefont {{A. J. Leggett,~{\etal}}}(1987)}]{Leggett}%
  \BibitemOpen
  \bibfield  {author} {\bibinfo {author} {\bibnamefont {{A. J.
  Leggett,~{\etal}}}},\ }\href {\doibase 10.1103/RevModPhys.59.1} {\bibfield
  {journal} {\bibinfo  {journal} {Rev. Mod. Phys.}\ }\textbf {\bibinfo {volume}
  {59}},\ \bibinfo {pages} {1} (\bibinfo {year} {1987})}\BibitemShut {NoStop}%
\bibitem [{\citenamefont {Zurek}(2003)}]{zurek}%
  \BibitemOpen
  \bibfield  {author} {\bibinfo {author} {\bibfnamefont {W.~H.}\ \bibnamefont
  {Zurek}},\ }\href {\doibase 10.1103/RevModPhys.75.715} {\bibfield  {journal}
  {\bibinfo  {journal} {Rev. Mod. Phys.}\ }\textbf {\bibinfo {volume} {75}},\
  \bibinfo {pages} {715} (\bibinfo {year} {2003})}\BibitemShut {NoStop}%
\bibitem [{\citenamefont {Bengtsson}\ and\ \citenamefont
  {{\.Z}yczkowski}(2006)}]{geometry}%
  \BibitemOpen
  \bibfield  {author} {\bibinfo {author} {\bibfnamefont {I.}~\bibnamefont
  {Bengtsson}}\ and\ \bibinfo {author} {\bibfnamefont {K.}~\bibnamefont
  {{\.Z}yczkowski}},\ }\href@noop {} {\emph {\bibinfo {title} {{Geometry of
  Quantum States: An Introduction to Quantum Entanglement}}}}\ (\bibinfo
  {publisher} {Cambridge University Press, Cambridge, U.K.},\ \bibinfo {year}
  {2006})\BibitemShut {NoStop}%
\bibitem [{\citenamefont {Gardas}(2009)}]{mgr}%
  \BibitemOpen
  \bibfield  {author} {\bibinfo {author} {\bibfnamefont {B.}~\bibnamefont
  {Gardas}},\ }\emph {\bibinfo {title} {Almost pure decoherence}},\ \href@noop
  {} {Master's thesis},\ \bibinfo  {school} {{I}nstitute of {P}hysics
  {U}niversity of {S}ilesia} (\bibinfo {year} {2009})\BibitemShut {NoStop}%
\bibitem [{\citenamefont {Gardas}(2010)}]{gardas}%
  \BibitemOpen
  \bibfield  {author} {\bibinfo {author} {\bibfnamefont {B.}~\bibnamefont
  {Gardas}},\ }\href {\doibase 10.1063/1.3442364} {\bibfield  {journal}
  {\bibinfo  {journal} {J. Math. Phys.}\ }\textbf {\bibinfo {volume} {51}},\
  \bibinfo {pages} {062103} (\bibinfo {year} {2010})}\BibitemShut {NoStop}%
\bibitem [{\citenamefont {Dajka}\ \emph {et~al.}(2009)\citenamefont {Dajka},
  \citenamefont {Mierzejewski},\ and\ \citenamefont {\L{}uczka}}]{dajka}%
  \BibitemOpen
  \bibfield  {author} {\bibinfo {author} {\bibfnamefont {J.}~\bibnamefont
  {Dajka}}, \bibinfo {author} {\bibfnamefont {M.}~\bibnamefont {Mierzejewski}},
  \ and\ \bibinfo {author} {\bibfnamefont {J.}~\bibnamefont {\L{}uczka}},\
  }\href {\doibase 10.1103/PhysRevA.79.012104} {\bibfield  {journal} {\bibinfo
  {journal} {Phys. Rev. A}\ }\textbf {\bibinfo {volume} {79}},\ \bibinfo
  {pages} {012104} (\bibinfo {year} {2009})}\BibitemShut {NoStop}%
\bibitem [{\citenamefont {Dajka}\ \emph {et~al.}(2008)\citenamefont {Dajka},
  \citenamefont {Mierzejewski},\ and\ \citenamefont {\L{}uczka}}]{dajka2}%
  \BibitemOpen
  \bibfield  {author} {\bibinfo {author} {\bibfnamefont {J.}~\bibnamefont
  {Dajka}}, \bibinfo {author} {\bibfnamefont {M.}~\bibnamefont {Mierzejewski}},
  \ and\ \bibinfo {author} {\bibfnamefont {J.}~\bibnamefont {\L{}uczka}},\
  }\href {\doibase 10.1103/PhysRevA.77.042316} {\bibfield  {journal} {\bibinfo
  {journal} {Phys. Rev. A}\ }\textbf {\bibinfo {volume} {77}},\ \bibinfo
  {pages} {042316} (\bibinfo {year} {2008})}\BibitemShut {NoStop}%
\bibitem [{\citenamefont {Alicki}\ and\ \citenamefont {Lendi}(1987)}]{alicki}%
  \BibitemOpen
  \bibfield  {author} {\bibinfo {author} {\bibfnamefont {R.}~\bibnamefont
  {Alicki}}\ and\ \bibinfo {author} {\bibfnamefont {K.}~\bibnamefont {Lendi}},\
  }\href@noop {} {\emph {\bibinfo {title} {Quantum dynamical semigroups and
  applications}}},\ Lecture notes in physics\ (\bibinfo  {publisher} {Springer,
  Berlin, Germany},\ \bibinfo {year} {1987})\BibitemShut {NoStop}%
\bibitem [{\citenamefont {\ifmmode \check{S}\else
  \v{S}\fi{}telmachovi\ifmmode~\check{c}\else \v{c}\fi{}}\ and\ \citenamefont
  {Bu\ifmmode~\check{z}\else \v{z}\fi{}ek}(2001)}]{korelacje}%
  \BibitemOpen
  \bibfield  {author} {\bibinfo {author} {\bibfnamefont {P.}~\bibnamefont
  {\ifmmode \check{S}\else \v{S}\fi{}telmachovi\ifmmode~\check{c}\else
  \v{c}\fi{}}}\ and\ \bibinfo {author} {\bibfnamefont {V.}~\bibnamefont
  {Bu\ifmmode~\check{z}\else \v{z}\fi{}ek}},\ }\href {\doibase
  10.1103/PhysRevA.64.062106} {\bibfield  {journal} {\bibinfo  {journal} {Phys.
  Rev. A}\ }\textbf {\bibinfo {volume} {64}},\ \bibinfo {pages} {062106}
  (\bibinfo {year} {2001})}\BibitemShut {NoStop}%
\bibitem [{\citenamefont {\ifmmode \check{S}\else
  \v{S}\fi{}telmachovi\ifmmode~\check{c}\else \v{c}\fi{}}\ and\ \citenamefont
  {Bu\ifmmode~\check{z}\else \v{z}\fi{}ek}(2003)}]{erratum}%
  \BibitemOpen
  \bibfield  {author} {\bibinfo {author} {\bibfnamefont {P.}~\bibnamefont
  {\ifmmode \check{S}\else \v{S}\fi{}telmachovi\ifmmode~\check{c}\else
  \v{c}\fi{}}}\ and\ \bibinfo {author} {\bibfnamefont {V.}~\bibnamefont
  {Bu\ifmmode~\check{z}\else \v{z}\fi{}ek}},\ }\href@noop {} {\bibfield
  {journal} {\bibinfo  {journal} {Phys. Rev. A}\ }\textbf {\bibinfo {volume}
  {67}} (\bibinfo {year} {2003})}\BibitemShut {NoStop}%
\bibitem [{\citenamefont {Hayashi}\ \emph {et~al.}(2003)\citenamefont
  {Hayashi}, \citenamefont {Kimura},\ and\ \citenamefont {Ota}}]{KrausRep}%
  \BibitemOpen
  \bibfield  {author} {\bibinfo {author} {\bibfnamefont {H.}~\bibnamefont
  {Hayashi}}, \bibinfo {author} {\bibfnamefont {G.}~\bibnamefont {Kimura}}, \
  and\ \bibinfo {author} {\bibfnamefont {Y.}~\bibnamefont {Ota}},\ }\href
  {\doibase 10.1103/PhysRevA.67.062109} {\bibfield  {journal} {\bibinfo
  {journal} {Phys. Rev. A}\ }\textbf {\bibinfo {volume} {67}},\ \bibinfo
  {pages} {062109} (\bibinfo {year} {2003})}\BibitemShut {NoStop}%
\bibitem [{\citenamefont {Langer}\ and\ \citenamefont {Tretter}(1998)}]{bom}%
  \BibitemOpen
  \bibfield  {author} {\bibinfo {author} {\bibfnamefont {H.}~\bibnamefont
  {Langer}}\ and\ \bibinfo {author} {\bibfnamefont {C.}~\bibnamefont
  {Tretter}},\ }\href@noop {} {\bibfield  {journal} {\bibinfo  {journal} {J.
  Operator Theory}\ }\textbf {\bibinfo {volume} {39}},\ \bibinfo {pages} {339}
  (\bibinfo {year} {1998})}\BibitemShut {NoStop}%
\bibitem [{\citenamefont {Adamjan}\ \emph {et~al.}(2001)\citenamefont
  {Adamjan}, \citenamefont {Langer},\ and\ \citenamefont {Tretter}}]{RiccEq}%
  \BibitemOpen
  \bibfield  {author} {\bibinfo {author} {\bibfnamefont {V.}~\bibnamefont
  {Adamjan}}, \bibinfo {author} {\bibfnamefont {H.}~\bibnamefont {Langer}}, \
  and\ \bibinfo {author} {\bibfnamefont {C.}~\bibnamefont {Tretter}},\ }\href
  {\doibase DOI: 10.1006/jfan.2000.3680} {\bibfield  {journal} {\bibinfo
  {journal} {Journal of Functional Analysis}\ }\textbf {\bibinfo {volume}
  {179}},\ \bibinfo {pages} {448 } (\bibinfo {year} {2001})}\BibitemShut
  {NoStop}%
\bibitem [{\citenamefont {Krovi}\ \emph {et~al.}(2007)\citenamefont {Krovi},
  \citenamefont {Oreshkov}, \citenamefont {Ryazanov},\ and\ \citenamefont
  {Lidar}}]{spinStar1}%
  \BibitemOpen
  \bibfield  {author} {\bibinfo {author} {\bibfnamefont {H.}~\bibnamefont
  {Krovi}}, \bibinfo {author} {\bibfnamefont {O.}~\bibnamefont {Oreshkov}},
  \bibinfo {author} {\bibfnamefont {M.}~\bibnamefont {Ryazanov}}, \ and\
  \bibinfo {author} {\bibfnamefont {D.~A.}\ \bibnamefont {Lidar}},\ }\href
  {\doibase 10.1103/PhysRevA.76.052117} {\bibfield  {journal} {\bibinfo
  {journal} {Phys. Rev. A}\ }\textbf {\bibinfo {volume} {76}},\ \bibinfo
  {pages} {052117} (\bibinfo {year} {2007})}\BibitemShut {NoStop}%
\bibitem [{\citenamefont {Arshed}\ \emph {et~al.}(2010)\citenamefont {Arshed},
  \citenamefont {Toor},\ and\ \citenamefont {Lidar}}]{spinStar2}%
  \BibitemOpen
  \bibfield  {author} {\bibinfo {author} {\bibfnamefont {N.}~\bibnamefont
  {Arshed}}, \bibinfo {author} {\bibfnamefont {A.~H.}\ \bibnamefont {Toor}}, \
  and\ \bibinfo {author} {\bibfnamefont {D.~A.}\ \bibnamefont {Lidar}},\ }\href
  {\doibase 10.1103/PhysRevA.81.062353} {\bibfield  {journal} {\bibinfo
  {journal} {Phys. Rev. A}\ }\textbf {\bibinfo {volume} {81}},\ \bibinfo
  {pages} {062353} (\bibinfo {year} {2010})}\BibitemShut {NoStop}%
\bibitem [{\citenamefont {Kittel}(1996)}]{Kittel}%
  \BibitemOpen
  \bibfield  {author} {\bibinfo {author} {\bibfnamefont {C.}~\bibnamefont
  {Kittel}},\ }\href@noop {} {\emph {\bibinfo {title} {Introduction to Solid
  States Physics}}}\ (\bibinfo  {publisher} {John Wiley \& Sons, Inc. NY},\
  \bibinfo {year} {1996})\BibitemShut {NoStop}%
\bibitem [{\citenamefont {\'Alvarez}\ and\ \citenamefont {Suter}(2010)}]{NMR}%
  \BibitemOpen
  \bibfield  {author} {\bibinfo {author} {\bibfnamefont {G.~A.}\ \bibnamefont
  {\'Alvarez}}\ and\ \bibinfo {author} {\bibfnamefont {D.}~\bibnamefont
  {Suter}},\ }\href {\doibase 10.1103/PhysRevLett.104.230403} {\bibfield
  {journal} {\bibinfo  {journal} {Phys. Rev. Lett.}\ }\textbf {\bibinfo
  {volume} {104}},\ \bibinfo {pages} {230403} (\bibinfo {year}
  {2010})}\BibitemShut {NoStop}%
\bibitem [{\citenamefont {{T. D. Ladd, F. Jelezko,~{\etal}}}(2010)}]{QCNature}%
  \BibitemOpen
  \bibfield  {author} {\bibinfo {author} {\bibnamefont {{T. D. Ladd, F.
  Jelezko,~{\etal}}}},\ }\href {\doibase 10.1038/nature08812} {\bibfield
  {journal} {\bibinfo  {journal} {Nature}\ }\textbf {\bibinfo {volume} {464}},\
  \bibinfo {pages} {45} (\bibinfo {year} {2010})}\BibitemShut {NoStop}%
\bibitem [{\citenamefont {Lvovsky}\ \emph {et~al.}(2009)\citenamefont
  {Lvovsky}, \citenamefont {Sanders},\ and\ \citenamefont {Tittel}}]{QMNature}%
  \BibitemOpen
  \bibfield  {author} {\bibinfo {author} {\bibfnamefont {A.~I.}\ \bibnamefont
  {Lvovsky}}, \bibinfo {author} {\bibfnamefont {B.~C.}\ \bibnamefont
  {Sanders}}, \ and\ \bibinfo {author} {\bibfnamefont {W.}~\bibnamefont
  {Tittel}},\ }\href {\doibase 10.1038/nphys1536} {\bibfield  {journal}
  {\bibinfo  {journal} {Nature Photonics}\ }\textbf {\bibinfo {volume} {3}},\
  \bibinfo {pages} {706} (\bibinfo {year} {2009})}\BibitemShut {NoStop}%
\bibitem [{\citenamefont {Reina}\ \emph {et~al.}(2002)\citenamefont {Reina},
  \citenamefont {Quiroga},\ and\ \citenamefont {Johnson}}]{Register2}%
  \BibitemOpen
  \bibfield  {author} {\bibinfo {author} {\bibfnamefont {J.~H.}\ \bibnamefont
  {Reina}}, \bibinfo {author} {\bibfnamefont {L.}~\bibnamefont {Quiroga}}, \
  and\ \bibinfo {author} {\bibfnamefont {N.~F.}\ \bibnamefont {Johnson}},\
  }\href {\doibase 10.1103/PhysRevA.65.032326} {\bibfield  {journal} {\bibinfo
  {journal} {Phys. Rev. A}\ }\textbf {\bibinfo {volume} {65}},\ \bibinfo
  {pages} {032326} (\bibinfo {year} {2002})}\BibitemShut {NoStop}%
\bibitem [{\citenamefont {{P. Neumann,~{\etal}}}(2010)}]{QRNature}%
  \BibitemOpen
  \bibfield  {author} {\bibinfo {author} {\bibnamefont {{P.
  Neumann,~{\etal}}}},\ }\href {\doibase 10.1038/nphys1536} {\bibfield
  {journal} {\bibinfo  {journal} {Nature Physics}\ }\textbf {\bibinfo {volume}
  {6}},\ \bibinfo {pages} {249} (\bibinfo {year} {2010})}\BibitemShut {NoStop}%
\bibitem [{\citenamefont {Cetinbas}\ and\ \citenamefont
  {Wilkie}(2008)}]{chaotic}%
  \BibitemOpen
  \bibfield  {author} {\bibinfo {author} {\bibfnamefont {M.}~\bibnamefont
  {Cetinbas}}\ and\ \bibinfo {author} {\bibfnamefont {J.}~\bibnamefont
  {Wilkie}},\ }\href {\doibase DOI: 10.1016/j.physleta.2007.09.047} {\bibfield
  {journal} {\bibinfo  {journal} {Physics Letters A}\ }\textbf {\bibinfo
  {volume} {372}},\ \bibinfo {pages} {1194 } (\bibinfo {year}
  {2008})}\BibitemShut {NoStop}%
\bibitem [{\citenamefont {Rosgen}(2008)}]{unital}%
  \BibitemOpen
  \bibfield  {author} {\bibinfo {author} {\bibfnamefont {B.}~\bibnamefont
  {Rosgen}},\ }\href {\doibase 10.1063/1.2992977} {\bibfield  {journal}
  {\bibinfo  {journal} {J. Math. Phys.}\ }\textbf {\bibinfo {volume} {49}},\
  \bibinfo {pages} {102} (\bibinfo {year} {2008})}\BibitemShut {NoStop}%
\bibitem [{\citenamefont {Fujii}\ and\ \citenamefont {Oike}(2010)}]{Rdiag}%
  \BibitemOpen
  \bibfield  {author} {\bibinfo {author} {\bibfnamefont {K.}~\bibnamefont
  {Fujii}}\ and\ \bibinfo {author} {\bibfnamefont {H.}~\bibnamefont {Oike}},\
  }\href@noop {} {\bibfield  {journal} {\bibinfo  {journal} {Int. J. Geom.
  Meth. Mod. Phys.}\ }\textbf {\bibinfo {volume} {7}} (\bibinfo {year}
  {2010})},\ \bibinfo {note} {accepted},\ \Eprint
  {http://arxiv.org/abs/1004.1207v2} {arXiv:1004.1207v2} \BibitemShut {NoStop}%
\bibitem [{\citenamefont {Marzlin}\ and\ \citenamefont
  {Sanders}(2004)}]{MSAdiab}%
  \BibitemOpen
  \bibfield  {author} {\bibinfo {author} {\bibfnamefont {K.-P.}\ \bibnamefont
  {Marzlin}}\ and\ \bibinfo {author} {\bibfnamefont {B.~C.}\ \bibnamefont
  {Sanders}},\ }\href {\doibase 10.1103/PhysRevLett.93.160408} {\bibfield
  {journal} {\bibinfo  {journal} {Phys. Rev. Lett.}\ }\textbf {\bibinfo
  {volume} {93}},\ \bibinfo {pages} {160408} (\bibinfo {year}
  {2004})}\BibitemShut {NoStop}%
\bibitem [{\citenamefont {{D. M. Tong,~\etal}}(2005)}]{TongAdiab}%
  \BibitemOpen
  \bibfield  {author} {\bibinfo {author} {\bibnamefont {{D. M. Tong,~\etal}}},\
  }\href {\doibase 10.1103/PhysRevLett.95.110407} {\bibfield  {journal}
  {\bibinfo  {journal} {Phys. Rev. Lett.}\ }\textbf {\bibinfo {volume} {95}},\
  \bibinfo {pages} {110407} (\bibinfo {year} {2005})}\BibitemShut {NoStop}%
\bibitem [{\citenamefont {Avron}\ and\ \citenamefont {Elgart}(1999)}]{Avron}%
  \BibitemOpen
  \bibfield  {author} {\bibinfo {author} {\bibfnamefont {J.}~\bibnamefont
  {Avron}}\ and\ \bibinfo {author} {\bibfnamefont {A.}~\bibnamefont {Elgart}},\
  }\href {\doibase 10.1007/s002200050620} {\bibfield  {journal} {\bibinfo
  {journal} {Commun. Math. Phys.}\ }\textbf {\bibinfo {volume} {203}},\
  \bibinfo {pages} {445} (\bibinfo {year} {1999})}\BibitemShut {NoStop}%
\bibitem [{\citenamefont {Jozsa}(1994)}]{fidelity}%
  \BibitemOpen
  \bibfield  {author} {\bibinfo {author} {\bibfnamefont {R.}~\bibnamefont
  {Jozsa}},\ }\href {\doibase 10.1080/09500349414552171} {\bibfield  {journal}
  {\bibinfo  {journal} {J. Mod. Opt.}\ }\textbf {\bibinfo {volume} {41}},\
  \bibinfo {pages} {2315} (\bibinfo {year} {1994})}\BibitemShut {NoStop}%
\bibitem [{\citenamefont {Pucha\l{}a}\ and\ \citenamefont
  {Miszczak}(2009)}]{JarekSF}%
  \BibitemOpen
  \bibfield  {author} {\bibinfo {author} {\bibfnamefont {Z.}~\bibnamefont
  {Pucha\l{}a}}\ and\ \bibinfo {author} {\bibfnamefont {J.}~\bibnamefont
  {Miszczak}},\ }\href {\doibase 10.1103/PhysRevA.79.024302} {\bibfield
  {journal} {\bibinfo  {journal} {Phys. Rev. A}\ }\textbf {\bibinfo {volume}
  {79}},\ \bibinfo {pages} {024302} (\bibinfo {year} {2009})}\BibitemShut
  {NoStop}%
\bibitem [{\citenamefont {{Z. Pucha\l{}a,~\etal}}(2009)}]{sub}%
  \BibitemOpen
  \bibfield  {author} {\bibinfo {author} {\bibnamefont {{Z.
  Pucha\l{}a,~\etal}}},\ }\href@noop {} {\bibfield  {journal} {\bibinfo
  {journal} {Quantum Inf. Comput.}\ }\textbf {\bibinfo {volume} {9}},\ \bibinfo
  {pages} {0103} (\bibinfo {year} {2009})}\BibitemShut {NoStop}%
\bibitem [{\citenamefont {Griffiths}(1995)}]{griff}%
  \BibitemOpen
  \bibfield  {author} {\bibinfo {author} {\bibfnamefont {D.}~\bibnamefont
  {Griffiths}},\ }\href@noop {} {\emph {\bibinfo {title} {Introduction to
  quantum mechanics}}}\ (\bibinfo  {publisher} {Prentice Hall, Inc.},\ \bibinfo
  {year} {1995})\BibitemShut {NoStop}%
\bibitem [{\citenamefont {{D. I. Schuster,~{\etal}}}(2007)}]{X}%
  \BibitemOpen
  \bibfield  {author} {\bibinfo {author} {\bibnamefont {{D. I.
  Schuster,~{\etal}}}},\ }\href {\doibase doi:10.1038/nature0546} {\bibfield
  {journal} {\bibinfo  {journal} {Nature}\ }\textbf {\bibinfo {volume} {445}},\
  \bibinfo {pages} {515} (\bibinfo {year} {2007})}\BibitemShut {NoStop}%
\bibitem [{\citenamefont {Sarandy}\ and\ \citenamefont {Lidar}(2005)}]{AA}%
  \BibitemOpen
  \bibfield  {author} {\bibinfo {author} {\bibfnamefont {M.~S.}\ \bibnamefont
  {Sarandy}}\ and\ \bibinfo {author} {\bibfnamefont {D.~A.}\ \bibnamefont
  {Lidar}},\ }\href {\doibase 10.1103/PhysRevA.71.012331} {\bibfield  {journal}
  {\bibinfo  {journal} {Phys. Rev. A}\ }\textbf {\bibinfo {volume} {71}},\
  \bibinfo {pages} {012331} (\bibinfo {year} {2005})}\BibitemShut {NoStop}%
\bibitem [{\citenamefont {Lidar}(2008)}]{AQQ}%
  \BibitemOpen
  \bibfield  {author} {\bibinfo {author} {\bibfnamefont {D.~A.}\ \bibnamefont
  {Lidar}},\ }\href {\doibase 10.1103/PhysRevLett.100.160506} {\bibfield
  {journal} {\bibinfo  {journal} {Phys. Rev. Lett.}\ }\textbf {\bibinfo
  {volume} {100}},\ \bibinfo {pages} {160506} (\bibinfo {year}
  {2008})}\BibitemShut {NoStop}%
\bibitem [{\citenamefont {Galindo}\ and\ \citenamefont
  {Pascual}(1990{\natexlab{b}})}]{galindo}%
  \BibitemOpen
  \bibfield  {author} {\bibinfo {author} {\bibfnamefont {A.}~\bibnamefont
  {Galindo}}\ and\ \bibinfo {author} {\bibfnamefont {P.}~\bibnamefont
  {Pascual}},\ }\href@noop {} {\emph {\bibinfo {title} {Quantum Mechanics vol
  1}}}\ (\bibinfo  {publisher} {Springer-Verlag, Berlin, Germany},\ \bibinfo
  {year} {1990})\BibitemShut {NoStop}%
\end{thebibliography}

             
  %

\end{document}